\documentclass[conference]{IEEEtran}

\usepackage{soul} 
\IEEEoverridecommandlockouts
\usepackage{cite}
\usepackage{amsmath,amssymb,amsfonts}
\usepackage{braket}
\usepackage{algorithmic}
\usepackage{multirow}
\usepackage{graphicx}
\usepackage{textcomp}
\usepackage{xcolor}
\def\BibTeX{{\rm B\kern-.05em{\sc i\kern-.025em b}\kern-.08em
    T\kern-.1667em\lower.7ex\hbox{E}\kern-.125emX}}
\begin{document}

\title{Randomized Reversible Gate-Based Obfuscation for Secured Compilation of Quantum Circuit\\
}

\author{\IEEEauthorblockN{Subrata Das}
\IEEEauthorblockA{\textit{Dept. of Electrical Engineering} \\
\textit{Pennsylvania State University}\\
University Park, PA \\
sjd6366@psu.edu}
\and
\IEEEauthorblockN{Swaroop Ghosh}
\IEEEauthorblockA{\textit{Dept. of Electrical Engineering} \\
\textit{Pennsylvania State University}\\
University Park, PA \\
szg212@psu.edu}

}

\maketitle

\begin{abstract}
The success of quantum circuits in providing reliable outcomes for a given problem depends on the gate count and depth in near-term noisy quantum computers. A circuit (that implements a given function) with a low gate count and short depth is more likely to give a correct solution than the circuit variant with a higher gate count and depth. As such, quantum circuit compilers that decompose high-level gates to native gates of the hardware and optimize the circuit play a key role in quantum computing. However, the quality and time complexity of the optimization process can vary significantly especially for practically relevant large-scale quantum circuits. As a result, third-party (often less-trusted/untrusted/unreliable) compilers have emerged, claiming to provide better and faster optimization of complex quantum circuits than so-called trusted compilers. However, untrusted compilers can pose severe security risks, such as the theft of sensitive intellectual property (IP) embedded within the quantum circuit. We propose an obfuscation technique for quantum circuits using randomized reversible gates to protect them from such attacks during compilation. The idea is to insert a small random circuit into the original circuit and send it to the untrusted compiler. Since the circuit function is corrupted, the adversary (i.e., untrusted compiler) may get incorrect IP. However, the user may also get incorrect output post-compilation. To circumvent this issue, we concatenate the inverse of the random circuit in the compiled circuit to recover the original functionality. We demonstrate the practicality of our method by conducting exhaustive experiments on a set of benchmark circuits and measuring the quality of obfuscation by calculating the Total Variation Distance (TVD) metric. Our method achieves TVD of up to 1.92 and performs at least 2X better than a previously reported obfuscation method. We also propose a novel adversarial reverse engineering (RE) approach and show that the proposed obfuscation is resilient  against RE attacks. The proposed technique introduces minimal degradation in fidelity ($\sim$1\% to $\sim$3\% on average).

\end{abstract}


\section{Introduction}
\label{sec:intro}
Quantum computing is a rapidly evolving field with potential to revolutionize various industries, including drug discovery, material science, and financial modeling \cite{bova2021commercial, national2019quantum}. At its core, quantum computing relies on the principles of quantum mechanics to perform computations \cite{divincenzo1995quantum, nielsen2002quantum}. The unique property of qubits is that they can exist in multiple states simultaneously, allowing quantum computers to solve certain complex problems exponentially faster than classical computers \cite{nielsen2002quantum}. This property is due to the ability of qubits to maintain quantum coherence, which is the ability of a quantum system to maintain a superposition state without collapsing into a definite state. The potential benefits of quantum computing have spurred significant research efforts worldwide, and several tech giants like IBM, Google, and Microsoft have already developed prototype quantum computers \cite{bova2021commercial, cusumano2018business}.

Despite the potential of quantum computing, the technology is still in its early stages, and several challenges must be addressed before it can be widely adopted \cite{national2019quantum, corcoles2019challenges}. One of the main challenges in quantum computing is the optimization of quantum circuits (i.e., an ordered sequence of quantum gates that perform specific operations on qubits to solve a problem) \cite{shaydulin2019hybrid}. 
Poorly optimized quantum circuits can produce random outcomes instead of the desired results due to noise and short coherence times. Therefore, optimizing quantum circuits has become an essential area of research in quantum computing. The optimization of quantum circuits is a complex task that requires specialized knowledge and tools. Several quantum compilers have been developed for this purpose including Qiskit, QuilC, and Forest \cite{anis2021qiskit, smith2020open}. These compilers translate high-level quantum circuit descriptions into low-level gates that are executed on quantum hardware. On one hand, the quality of the optimization and the compilation time can vary significantly especially for large-scale quantum circuits. On the other hand, third-party compilers (less-trusted/untrusted/unreliable) have also emerged, claiming to provide better and faster optimization of complex quantum circuits than established (trusted) compilers \cite{zapatacomputing_2023, pytket}. However, using untrusted compilers to optimize quantum circuits can pose severe security risks, such as the theft of sensitive intellectual property (IP). Due to the vast potential of this technology, many organizations are investing heavily in research and development to gain a competitive edge. The theft of sensitive quantum circuit designs or algorithms can provide an unfair advantage to competitors or malicious actors, allowing them to develop similar technology without investing the same level of time and resources. This can result in significant financial losses for the organization that originally developed the IP and can also slow down the pace of quantum computing.
Therefore, novel techniques are needed to optimize quantum circuits while preserving their confidentiality. The objective of this paper is to propose an obfuscation approach to protect the IP of quantum circuits during optimization by untrusted third-party compilers. 

\subsection{Proposed Idea}
\label{subsec:proposed_idea}


The proposed method involves inserting a random quantum circuit into the original vulnerable circuit before sending it to the untrusted compiler. The gates within the random circuit can be chosen by the user either randomly or following any design technique to ensure maximal corruption of the true functionality. 
In order to extract the original circuit, the adversary will have to identify and remove this random circuit from the obfuscated circuit which is computationally quite difficult. Any gate in the obfuscated design can be a part of the random circuit which makes it hard for the adversary to brute-force. Moreover, the adversary cannot validate his/her brute-force guess due to the lack of a golden/oracle model. If the adversary tries to reuse the obfuscated circuit without removing the random circuit, he/she will get a corrupted result or severely degraded performance. Thus, the insertion of a random circuit will protect the IP by hiding the true functionality of the original circuit from the rouge adversary. 

\begin{figure}[t]
    \includegraphics[width=0.5\textwidth]{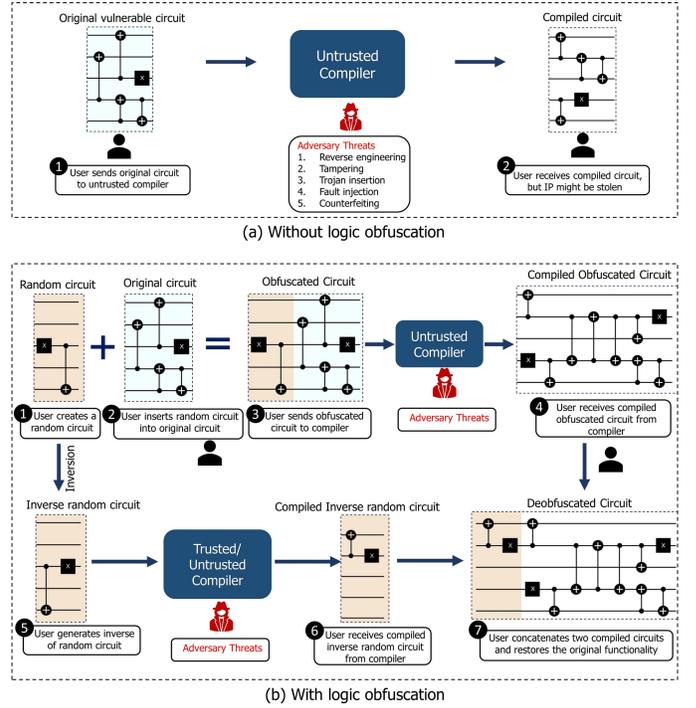}
    \centering
\caption{(a) Proposed attack model. The user sends the original quantum circuit to the untrusted compiler, where the adversary can steal the IP or RE the circuit. (b) Proposed logic obfuscation technique as a countermeasure. The user creates a random circuit (1), inserts it into the original circuit (2), and gets the obfuscated circuit. Then the user sends the obfuscated circuit to the untrusted compiler (3) and receives it back after compilation (4). In the meantime, the user also generates the inverse of the random circuit (5) and compiles it separately using a trusted/untrusted compiler (6). Finally, the user concatenates the two compiled circuits and restores the original functionality (7). }
    \label{fig:schematic}
    \vspace{-.31cm}
\end{figure}

One of the major limitations of existing quantum circuit obfuscation techniques \cite{suresh2021short} is that they require some sort of identifier/marker (e.g., barrier) at the location of fake gate insertion, which in turn, may provide a clue to the adversary. The marker is required so that the user can remove the fake gate post compilation. We overcome the above challenge by exploiting the reversibility property of quantum computing. In the proposed approach, the user will generate an inverse quantum circuit of the random circuit (shallow in general) using a trusted compiler and insert at the random circuit insertion point to restore the original functionality after compilation. Since the random and inverse random circuits consist of few gates, the compilation quality will be reasonable with trusted compilers. 
Fig. \ref{fig:schematic} illustrates this idea further.

\subsection{Our Contributions}\label{subsec:our_contributions} We (a) propose a novel quantum circuit obfuscation methodology that uses random reversible quantum gates to corrupt the original functionality, (b) evaluate the performance of the proposed method by conducting exhaustive experiments on a set of benchmark circuits and comparing them with previously reported obfuscation technique, (c) develop intuitions from the results and devise a technique to refine the random circuit for maximal obfuscation, (d) propose a deobfuscation procedure based on the concept of quantum reversibility, (e) propose a novel reverse engineering approach and evaluate its effectiveness against the proposed obfuscation.


\subsection{Paper Organization}
\label{subsec:paper_organization}

The remainder of this paper, Section \ref{sec:background} provides an overview of quantum computing and two metrics i.e., Total Variation Distance (TVD) and Degree of Functional Corruption (DFC), which we have used to quantify the quality of functional obfuscation. We also review the related work. Section \ref{sec:threat_model} outlines the threat model and the adversarial  capabilities. Section \ref{sec:proposed_technique} presents the proposed obfuscation technique and a case study demonstrating its effectiveness. Section \ref{sec:results_analysis} shows the experimental results and analysis of the proposed technique. Section \ref{sec:deobfus} describes the post-compilation deobfuscation methodology and detailed overhead analysis. Section \ref{sec:reverse} presents a novel reverse engineering effort analysis method against the proposed obfuscation technique. 
Finally, Section \ref{sec:conclusion} concludes the paper.

\section{Background} 
\label{sec:background}
In this section, we provide a brief introduction to the fundamental concepts of quantum computing that are relevant to our work.
\subsection{Quantum Computation Preliminaries}
\subsubsection{Qubits} Qubits are the fundamental building blocks of quantum information. Unlike classical bits that can exist in only two states (0 or 1), a qubit can exist in a superposition of both states at the same time. This superposition is a key characteristic of quantum computing, as it allows for multiple calculations to be performed simultaneously. 
The most common representation of a qubit is a two-level system, with the basis states $\ket{0}$ and $\ket{1}$. The state of a single qubit can be represented as a linear combination of these two states, denoted by $\ket{\Psi}=\alpha\ket{0}+\beta\ket{1}$, where $\alpha$ and $\beta$ are complex numbers and the squared magnitudes of $\alpha$ and $\beta$ represent the probabilities of measuring the qubit in the states $\ket{0}$ and $\ket{1}$, respectively \cite{nielsen2002quantum}. 

\subsubsection{Quantum Gates} 
Quantum gates operate on qubits to perform specific operations such as, changing the state of a qubit, entangling multiple qubits, and creating superposition states. Each quantum gate is represented by a unitary matrix which describes the transformation it performs on the quantum state. Some commonly used quantum gates include the Pauli-X gate, Hadamard gate, and controlled-NOT (CNOT) gate. The Pauli-X gate is used to flip the state of a qubit, 
while the CNOT gate is a two-qubit gate that applies the NOT operation to the second qubit if the first qubit is in the state $\ket{1}$. Mathematically, a quantum gate is represented by a unitary matrix U, which satisfies the condition $U^{\dagger}$$U=I$, where $U^{\dagger}$ denotes the conjugate transpose of U and $I$ denotes the identity matrix. The action of a quantum gate on a qubit state $\ket{\psi}$ is given by $U\ket{\psi}$. This operation preserves the norm of the state vector and can be interpreted as a rotation in the Bloch sphere representation of the qubit state.

\subsubsection{Quantum Reversibility} Quantum computation is a reversible process, meaning that information can be retrieved from the output and the computation can be undone by applying the same operations in reverse order. This property is essential in quantum computing and is a consequence of the fundamental unitary evolution of quantum systems. Mathematically, a quantum circuit is represented by a sequence of unitary transformations, or quantum gates, that act on qubits. A unitary transformation U is reversible since it can be undone by applying the inverse transformation $U^{\dagger}$. For example, the Hadamard gate H is a unitary transformation that maps the basis state $\ket{0}$ to the superposition state $(\ket{0}+\ket{1})/\sqrt{2}$, and maps the basis state $\ket{1}$ to the superposition state $(\ket{0}-\ket{1})/\sqrt{2}$. Applying the Hadamard gate again to these superposition states brings them back to the original basis states.
In practice, the reversibility of quantum circuits is implemented by applying the inverse of each gate in reverse order. This can be achieved by using the adjoint matrix of the gate, which is defined as the conjugate transpose of the gate matrix. For example, if a quantum circuit applies a gate $U$ followed by a gate $V$, the reverse circuit would apply the gate $V^{\dagger}$ followed by the gate $U^{\dagger}$.

\subsubsection{Compilation of Quantum Circuits} Compilation refers to the process of converting a high-level quantum circuit representation into an executable form that can be run on a physical quantum computer. It is a critical step in quantum computing as it translates abstract circuits into physical implementations that are compatible with the hardware constraints. The compilation process in a typical quantum compiler such as, Qiskit consists of the following steps:

\begin{itemize}
    
\item \textit{Virtual circuit optimization.}  In this step, the compiler applies a set of optimization techniques to the high-level quantum circuit to reduce the overall number of gates and improve the circuit's performance. For example, the compiler can use techniques such as, gate cancellation or circuit simplification to remove redundant gates or simplify the circuit's structure.
\item \textit{3-qubit gate decomposition.} Current quantum computers have limitations in the number of qubits and the types of gates they support. Therefore, a high-level quantum circuit that contains multi-qubit gates may need to be decomposed into a series of single- and two-qubit gates that can be executed on the available hardware. For example, a 3-qubit Toffoli gate can be decomposed into 6 CNOT gates and several single-qubit gates.
\item \textit{Placement on physical qubits.} In this step, the logical qubits of the circuit are mapped onto the physical qubits of the target device. This process is necessary because quantum hardware usually has limited connectivity between the qubits, meaning that not all qubits can be directly connected. 
The coupling constraint (represented using a coupling map) of a quantum device describes the set of allowed two-qubit interactions that can be performed on the device. 
The coupling map can be represented as a graph, where the vertices represent the qubits and the edges represent the allowed two-qubit interactions \cite{wille2019mapping}.
The qubit mapping process aims to find a mapping that satisfies the coupling constraint of the target device while optimizing some other criteria such as, gate count, circuit depth, or error rate. The qubit mapping problem is known to be NP-hard, and several heuristic algorithms have been proposed to solve it \cite{wille2019mapping, bhattacharjee2019muqut}.\\
For example, let us consider a quantum circuit with two qubits, $q_{0}$ and $q_{1}$, where a CNOT gate is applied on $q_{0}$ and $q_{1}$. If the device's coupling constraint only allows CNOT gates between adjacent qubits, then we need to map $q_{0}$ and $q_{1}$ to adjacent physical qubits on the device. The starting physical-to-virtual (p2v) qubit mapping is known as the \emph{initial layout}. If $q_{0}$ and $q_{1}$ are not adjacent in the initial layout, we can use swap gates to move $q_{0}$ or $q_{1}$ to an adjacent qubit, perform the CNOT gate, and then move it back to its original position. The \emph{final layout} represents the resolved coupling constraints and may differ from the initial layout due to the addition of SWAP gates. This mapping process introduces additional gates and increases the circuit depth, so the mapping should be optimized to minimize the overhead.

\item \textit{Routing.} After placement, the compiler needs to determine the path that each qubit will take to execute the circuit. This process is known as routing. A routing algorithm must take the coupling constraints into account to minimize the number of swap gates and optimize the circuit's execution time.

\item \textit{Translation to Basis Gates.} Quantum computers support a restricted set of gates known as basis gates, which are usually single-qubit rotations and two-qubit entangling gates. Any quantum circuit can be implemented using a combination of basis gates. Therefore, the next step in the compilation process is to translate the circuit into the basis gates supported by the target device. For example, the IBM Q systems support the basis gates {‘ID’, ‘SX’, ‘X’, ‘CX’}.

\item  \textit{Physical Circuit Optimization.} After translating the circuit to the basis gates, the compiler can optimize the physical implementation of the circuit on the target device. Physical optimization includes techniques such as, gate fusion, which involves combining multiple gates into a single gate, and gate cancellation, which involves removing consecutive gates that cancel each other out. The goal of physical optimization is to reduce the total number of gates and the overall circuit execution time. For instance, consider a quantum circuit that applies two consecutive X gates on the same qubit. The first X gate flips the qubit from $\ket{0}$ and $\ket{1}$, and the second X gate flips it back to $\ket{0}$. These two gates can be canceled out. 
\end{itemize}

\subsection{Metrics to Quantify Obfuscation Quality}

\subsubsection{Total Variation Distance (TVD)} TVD is a statistical metric used to quantify the difference between two probability distributions. In this study, we use TVD to measure the degradation in the output of an obfuscated quantum circuit compared to the true output. The TVD is calculated as the sum of the absolute differences between the counts of each element in the output distribution of the obfuscated circuit and the output distribution of the original circuit, divided by the total number of shots. TVD can be represented by,
\begin{equation}
    TVD = \frac{\sum_{i}^{}(\left|x_{i,orig}-x_{i,obfus}\right|)}{Total\;no.\;of\;shots}
\end{equation}

Here, $x_{i,orig}$ and $x_{i,obfus}$ represent the count of the $i^{th}$ element in the output distribution of the original circuit and the obfuscated circuit, respectively. The resulting TVD value provides a measure of the distance between the two distributions, with a value of 0 indicating that the distributions are identical and a larger value indicating a greater difference between the distributions. Therefore, a higher TVD value indicates stronger obfuscation. While TVD is a useful metric for quantifying the difference between two probability distributions, it may not capture change in only one element of the distribution, namely the correct basis state probability. To demonstrate this, let us consider the output of a quantum circuit for two possibilities, such as, \{“0":95, “1":5\}, which is the correct output over 100 shots. Subsequently, we insert a random circuit into this circuit and obtain an output of \{“0":55, “1":45\} where the correct output “0" is obtained for the majority of shots even after obfuscation, indicating that the random circuit did not completely obscure the original functionality.  However, calculating the TVD results in a high value of 0.8, indicating satisfactory obfuscation. Thus, the TVD alone cannot express the degree of functional corruption comprehensively making it important to consider other metrics that may better capture this type of corruption.



\subsubsection{Degree of Functional Corruption (DFC)} We propose an additional new metric, the Degree of Functional Corruption (DFC), to evaluate the obfuscated circuit's output. DFC is defined as the difference between the count of the correct output of the obfuscated circuit and the count of the maximum incorrect output of the obfuscated circuit, normalized by the total number of shots. 
DFC varies between -1 and 1, where a negative value indicates that the obfuscated circuit produces the wrong output, and a positive value indicates it produces the correct output for the majority of shots even after inserting the random circuit. For example, in the aforementioned case, DFC is +0.1, indicating that the functionality was not fully corrupted by the random circuit. Therefore, a lower DFC value indicates better concealing of the original circuit's true functionality which is desired.
\begin{figure*}[t]
    \includegraphics[width=0.8\textwidth]{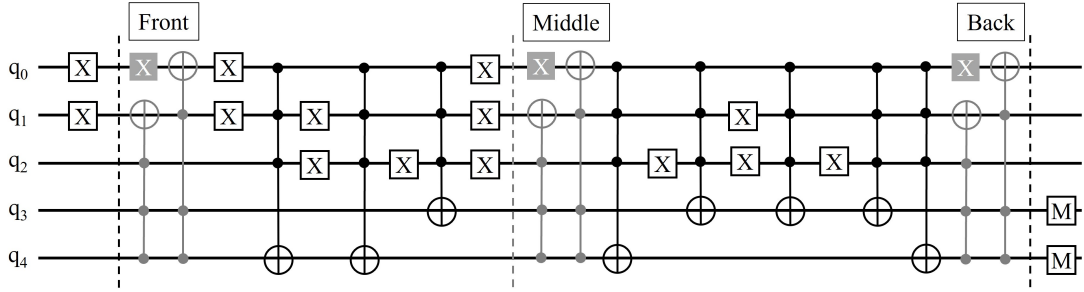}
    \centering
    \caption{Circuit diagram of a 1-bit adder with random circuit inserted at front, middle, and back position. Vertical dashed lines indicate barriers. Black gates and barriers belong to the original circuit, and the gray barrier and gates are part of the random circuit. }
    \label{fig:adder}
    \vspace{-.31cm}
\end{figure*}

\subsection{Related Work and Their Limitations} 

The attack model addressed in this paper is the same as the one proposed in \cite{suresh2021short}. The authors propose inserting a single dummy CX gate into the quantum circuit to corrupt the functionality (for obfuscation) before sending it to an untrusted compiler. After compilation, the user needs to discover and delete the dummy gate to restore functionality which can be difficult due to the optimization. Therefore, this technique requires adding barriers before and after the dummy CX gate to prevent optimization across the barriers and to make it recognizable after compilation. However, these added barriers will not only impair the quality of optimization but also leave an obvious clue for the adversary.

Another related work \cite{saki2021split} proposes a split compilation technique to secure quantum IPs from untrusted compilers. The idea is to split the quantum circuit into multiple parts that are sent to a single untrusted compiler at different times or to multiple untrusted compilers simultaneously, providing partial information to adversaries. The authors show that the split compilation can secure IPs and introduce factorial time reconstruction complexity. However, it is important to note that the approach does incur an overhead of 3\%-6\% on average which may limit its practical use in certain scenarios. Furthermore, the technique can face challenges under colluding adversaries who can recover the full circuit. If the split circuits are sent to the same compiler at different times, there is also a likelihood of full netlist recovery if the adversary can match the user and the workload. Our work aims to address these limitations by proposing a complementary methodology that ensures security while minimizing the overhead. The proposed approach can be employed alongside split compilation for an additional layer of security.


\section{Threat model and adversary capability} 
\label{sec:threat_model}
\subsection{Threat Model} 
In this work, we consider the quantum circuits as valuable IP since it requires substantial investment in time and resources to create them and they may include sensitive information. We also assume that the user may employ untrusted/less-trusted third-party compilers as proposed in earlier works \cite{zapatacomputing_2023,pytket}. There are various reasons why a quantum circuit designer might choose to use a third-party compilation service that they cannot fully trust. These include the importance of optimizing the circuit to achieve meaningful results with NISQ computers and the scarcity of trusted compilers that have kept pace with the latest advancements in optimization. Nevertheless, utilizing an untrusted compiler presents a significant risk of IP theft. In addition to the risk of IP theft, the quantum circuits sent to untrusted compilers can be subjected to impactful Trojan insertion and/or tampering such as, addition or removal of quantum gates. If the circuit is sent in clear, an adversary may be able to modify the circuit in a way that is not easily detectable, compromising the integrity of the circuit and the computation outcome. This is particularly concerning when the circuit contains sensitive information or is used in critical applications such as, financial analysis or national security.


\subsection{Adversary Capability} 
We make the following assumptions regarding the compiler package used for quantum circuits: (a) the package is hosted remotely by an untrusted third party and some rogue adversary can attack the netlist to retrieve sensitive information from the unoptimized quantum circuit, (b) the adversary may assume that the circuit is obfuscated and might also have full knowledge of the obfuscation algorithm, possibly through leaked information from the design house. However, the adversary does not have access to an oracle model to validate his/her assumption and, (c) the adversary has sufficient computational resources to scrutinize the circuit and detect clues such as, barriers inserted in the circuit to identify the random circuit. The goal of the adversary is to recover the functionality, algorithm, or any other confidential information that may be encoded in the unoptimized quantum circuit.

\section{Proposed Obfuscation Technique} 
\label{sec:proposed_technique}
In this section, we provide a brief overview of the proposed obfuscation technique and demonstrate its effectiveness using a case study. Next, we analyze the results of the example circuit and extract salient features of the random circuit which contribute to higher obfuscation. Based on the intuitions developed, we further refine the random circuit for maximal functionality corruption. 
\subsection{Overview}

The idea of the proposed technique is to insert a random circuit into the original circuit to conceal its functionality from the untrusted compiler. The user can choose the type and depth of the random circuit to make it compatible with the original circuit and ensure that minimal clues are left for the rogue adversary. However, complete obfuscation is challenging because of the probabilistic nature of quantum computation and the inherent margin between correct/incorrect outputs. For instance, if a single qubit quantum circuit has the probability of generating correct output (e.g. 0) in 95\% measurements and incorrect output (e.g. 1) in 5\% measurements, then the logic obfuscation technique must overcome this margin of 90\% to corrupt the functionality fully and alter the result. Hence, the insertion of the random circuit should be done strategically to ensure higher obfuscation. To address this issue, we further propose a design technique to slightly refine the randomly generated circuit to enhance functionality corruption. The insertion location of the random circuit is another crucial aspect of the proposed technique as it directly impacts the effectiveness of obfuscation as well as deobfuscation by the user post-compilation. Strategic placement of the random circuit can enhance functionality corruption, minimize clues for potential adversaries, and mitigate the risk of detection by the untrusted compiler. In this work, we consider three insertion locations namely, front of the circuit, back of the circuit and middle of the circuit.



\subsection{Case study: 1-bit Adder}

To illustrate the idea, we apply the proposed obfuscation technique to a 5-qubit benchmark circuit i.e. 1-bit adder as shown in Fig. \ref{fig:adder}. This circuit contains twenty $X$ and $C3X$ ( 3-qubit controlled X gate) quantum gates and can add 3 inputs at qubits $q_{0}$, $q_{1}$ and $q_{2}$. Outputs i.e. carry and sum are measured at qubits $q_{3}$, $q_{4}$, respectively. At first, the original 1-bit adder circuit is simulated using the noise model of the $ibmq\_valencia$ device with 10,000 shots. We use the Qiskit from IBM for our simulations. Next, we generate 100 random circuits containing $X$, $CX$, and $C3X$ using Qiskit for obfuscation and insert one random circuit at a time in front of the original circuit (gray gates in Fig. \ref{fig:adder} indicate the random circuit) and re-simulate the obfuscated circuit. Further, we insert the same 100 random circuits one at a time at the middle and back of the original circuit and run the same experiments. Notably, for inserting the random circuit in the middle, we add an extra barrier beside the random circuit so that the user can identify the insertion location after compilation which is necessary for successful deobfuscation (will be further discussed in Section \ref{sec:deobfus}). For all the experiments, we measure the quality of obfuscation by calculating the TVD and DFC metrics as described in Section \ref{sec:background}. 

\begin{figure}[t]
     \includegraphics[width=0.3\textwidth]{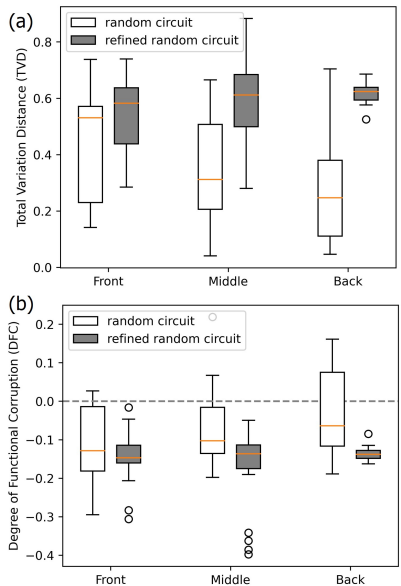}
    \centering
    \caption{Distribution of (a) TVD and (b) DFC values for inserting 100 random circuits and 100 refined random circuits one at a time at the front, middle, and back of the 1-bit adder circuit.}
    \label{fig:adder_results}
\end{figure}

Fig. \ref{fig:adder_results} shows the distribution of these TVD and DFC values for 100 random circuits and three insertion locations i.e. front, middle, and back. From the white box plots in Fig. \ref{fig:adder_results}(a), we can see that inserting random circuits at the front yields a median TVD of around 0.57 which is slightly higher than the middle and back locations. Noticeably, all three cases have wide TVD distributions with values ranging from 0.74 to 0.04. Such observation indicates that the TVD might vary over a wide range if we choose the extra quantum gates totally randomly. Fig. \ref{fig:adder_results}(b) provides further evidence for this idea demonstrating that we get positive DFC values (as high as 0.22) for some random circuits. According to the definition, positive DFC values indicate partial corruption of the original functionality which is not desired for high-quality obfuscation. Therefore, we analyze the 100 random circuits and identify the features that play key role in deciding the quality of obfuscation. Interestingly, we note that the random circuits with an X gate on any of the measurement qubits i.e., $q_{3}$ or $q_{4}$ always yield a negative DFC value and significantly higher TVD compared to the other random circuits. This can be attributed to the fact that the X gate of the random circuit flips the measurement qubit resulting in complete functional corruption. Based on this intuition, we develop a simple technique to refine the randomly generated circuit for enhancing the quality of obfuscation which will be discussed in the next section.

\subsection{Random Circuit Refinement Technique }

\begin{figure}[t]
    \includegraphics[width=0.5\textwidth]{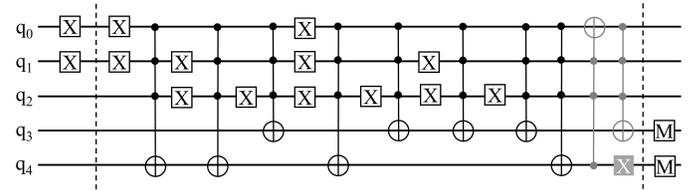}
    \centering
    \caption{A refined random circuit is inserted at the back of a 1-bit adder circuit. Black and gray gates represent the original and random circuits, respectively.}
    \label{fig:refined_circuit}
    \vspace{-0.3cm}
\end{figure}

Our proposed random circuit refinement technique involves inserting an X gate on any of the measurement qubits. The rest of the extra gates are generated and placed randomly only on the remaining qubits to ensure that the measurement qubit is not flipped again. Fig. \ref{fig:refined_circuit} illustrates an obfuscated design using a refined random circuit in which an X gate is placed on one of the measurement qubit $q_{4}$ and other random gates are placed on the rest of the qubits. To demonstrate the effectiveness of this technique, we generate 100 such refined random circuits and insert one refined random circuit at a time at the front, middle, and back of the 1-bit adder circuit. We conduct the simulation under the same experimental conditions as mentioned earlier and present the result in Fig. \ref{fig:adder_results}. The gray box plots in Fig. \ref{fig:adder_results}(a) show that the refined random circuits shorten the variance of the TVD distribution as well as increase the median TVD for all three insertion locations. Especially, the improvement is significant for the cases in which we add refined random circuits at the middle and back of the original circuit. Fig. \ref{fig:adder_results}(b) further shows that we can achieve negative DFC values i.e., complete functional corruption for all the 100 refined random circuits, and thereby, confirms the reliability of the refined obfuscation procedure.

\section{Experimental results and analysis}
\label{sec:results_analysis}

\subsection{Simulation Setup}
We use the IBM Qiskit for the simulations \cite{anis2021qiskit} which runs locally on an AMD Ryzen 7 4800U CPU (1.80 GHz) machine with 16 GB RAM (Windows 11 Pro). We take 10 benchmark circuits from the RevLib repository \cite{wille2008revlib} which are commonly used in contemporary works on quantum circuit compilations. The benchmark circuits contain various gate operations, various numbers of gates namely, 6, 8, 12, 14, 18, 20, 28, 36, 80 and 111, and various qubit sizes namely, 4, 5, 7, 10, and 12. The simulations are done for 10,000 shots. For simulating the smaller circuits (with up to 20 quantum gates), we use the $FakeValencia$ backend of IBM Qiskit which uses the realistic noise model of $ibmq\_valencia$ device. For the larger circuits, we use the noisy $QasmSimulator$ backend with our customized noise model. We tune the noise values of $ibmq\_valencia$ device in a way such that the noise does not corrupt the true functionality of the original circuit and we can assess the effectiveness of obfuscation. We generate the random circuits using Qiskit. To ensure that no clues are left for potential adversaries, the type of quantum gates used in the random circuit is chosen to match the gate types present in the original circuit. For example, as illustrated in Fig. \ref{fig:refined_circuit}, we have specifically chosen $X$ and $C3X$ gates to create a random circuit that are inserted into the 1-bit adder circuit for obfuscation. This choice is made based on the fact that the original adder circuit only contains these two types of gates. By using gate types that are already present in the original circuit, the random circuit can blend in seamlessly, making it harder for adversaries to discern any clues. 


\subsection{Obfuscation Quality Analysis}

\begin{figure}[t]
    \includegraphics[width=0.45\textwidth]{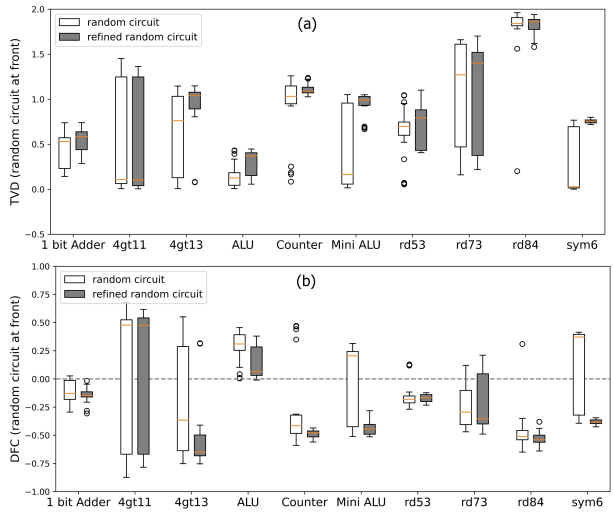}
    \centering
    \caption{{Distribution of (a) TVD and (b) DFC values for inserting 100 random circuits and 100 refined random circuits one at a time at the front of 10 benchmark circuits.} }
    \label{fig:benchmark_front}
    \vspace{-.31cm}
\end{figure}
\begin{figure}[b]
    \includegraphics[width=0.45\textwidth]{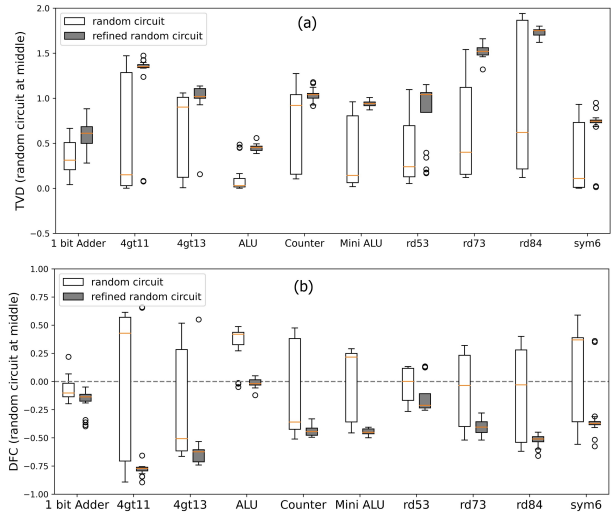}
    \centering
    \caption{Distribution of (a) TVD and (b) DFC values for inserting 100 random circuits and 100 refined random circuits one at a time in the middle of 10 benchmark circuits. }
    \label{fig:benchmark_middle}
    \vspace{-.31cm}
\end{figure}
\begin{figure}[t]
    \includegraphics[width=0.45\textwidth]{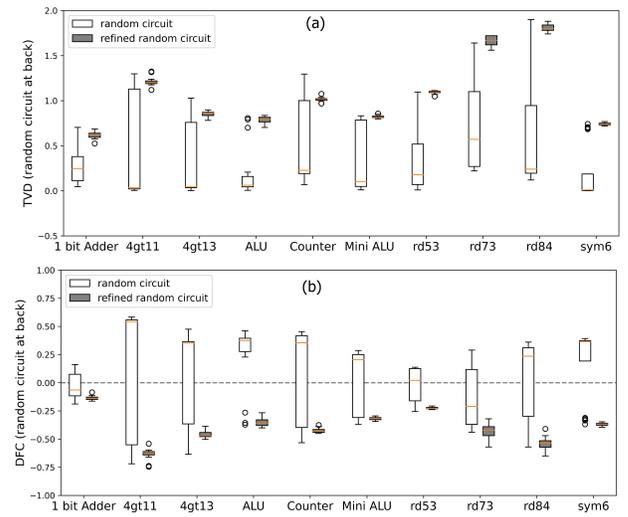}
    \centering
    \caption{Distribution of (a) TVD and (b) DFC values for inserting 100 random circuits and 100 refined random circuits one at a time at the back of 10 benchmark circuits. }
    \label{fig:benchmark_back}
    \vspace{-.31cm}
\end{figure}

Fig. \ref{fig:benchmark_front} demonstrates the distribution of TVD and DFC values for inserting 100 random and 100 refined random circuits one at a time at the front of 10 benchmark circuits. It can be inferred that when we insert totally random circuit at the front of the original benchmark circuits, we get a large variance of TVD and DFC distribution with some undesirable positive DFC values for 6 benchmarks (i.e., 4gt11, 4gt13, ALU, Mini ALU, rd73, sym6). The proposed refined random circuit-based obfuscation technique reduces the variance of distribution and increases the mean TVD for all the benchmark circuits. We calculate an average mean-TVD increase of around 57\% after applying the refinement technique. However, we can observe that 3 benchmark circuits, i.e., 4gt11, ALU and rd73 still show positive DFC values. Such observation indicates that depending on the design of the original circuit, some obfuscated designs might generate correct output even after the insertion of the refined random circuit at the front.\\
Next, we insert the same 100 random and 100 refined random circuits in the middle of 10 benchmark circuits. The results of TVD and DFC distributions are presented in Fig. \ref{fig:benchmark_middle}. Notably, the variance of TVD and DFC distributions reduce considerably and the mean TVD values rise significantly for all benchmarks with refined random circuits. The average mean-TVD increase is around 119\% indicating a strong logical obfuscation. Moreover, we get negative DFC values for all benchmark circuits (except for a very few cases in ALU) which further confirms that adding the refined random circuit in the middle results in higher functional corruption as compared to adding it at the front.\\
 Finally, the same 100 random and 100 refined random circuits are added at the back of 10 benchmark circuits and the results are presented in Fig. \ref{fig:benchmark_back}. Interestingly, for all benchmarks, the variance of TVD and DFC distributions reduces with very few outliers. As compared to the totally random circuits, we get an average mean-TVD increase of around 190\% using the refined random circuits. We also get negative DFC values for all the benchmarks (including ALU) in this case which is an indication of complete functional corruption.\\
\vspace{-.3cm}
\subsection{Comparative Analysis with Previous Work}
\label{sec:previous}
\begin{figure}[!t]
    \includegraphics[width=0.45\textwidth]{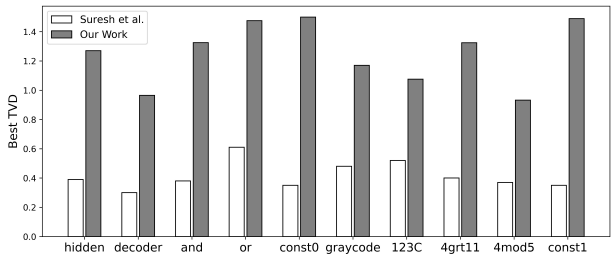}
    \centering
    \caption{Comparison between the best TVDs reported by Suresh \textit{et al.}\cite{suresh2021short} and the best TVDs obtained by our proposed obfuscation technique for 10 benchmark circuits. }
    \label{fig:comparison}
    \vspace{-.31cm}
\end{figure}

We compared the results of our proposed obfuscation technique with a closely related prior work \cite{suresh2021short}. For a systematic comparison, we have studied the same benchmark circuits which were used in \cite{suresh2021short}. We applied the proposed obfuscation technique by inserting 100 refined random circuits once at a time at the back of each original benchmark circuit and calculated the best i.e., highest TVD. Fig. \ref{fig:comparison} shows the results of this comparative analysis where the white bars represent the best TVD from \cite{suresh2021short} and the gray bars denote the best TVD obtained using the proposed obfuscation. The results indicate that our  obfuscation technique consistently outperforms the obfuscation methodology proposed in \cite{suresh2021short} for all ten benchmarks. Specifically, our TVD is at least 2X higher than the best TVD reported by \cite{suresh2021short} for each benchmark circuit. For some circuits such as, \textit{const1}, \textit{const0} and \textit{hidden}, our TVD is more than 4X higher, indicating superior obfuscation compared to \cite{suresh2021short}.

\subsection{Summary of Findings}
Based on the outcomes stated above, we can infer that inserting the refined random circuits closer to the output leads to higher logical obfuscation and stronger functional corruption. This can be ascribed to the fact that when we insert the random circuit far away from the output, their effects might get canceled out due to the operations of other quantum gates present in the circuit. On the contrary, when the random circuit is inserted closer to the output, their effect is immediately reflected in the result because of which we can achieve maximal functional corruption in such cases. 
\subsection{Potential Limitations}
One limitation of the proposed random circuit refinement technique, which inserts an \textit{X} gate on the measurement qubit, is that it can leave a clue for the adversary if the original circuit does not contain any \textit{X} gate. Therefore, it is advisable to avoid usage of the proposed refinement technique and only stick to inserting random circuits if the circuit do not contain any \textit{X} gate. However, relying solely on random circuits may result in weaker obfuscation which may not be desirable.


Another potential challenge is to restrict the number of gates and depth of the random circuit to ensure that the obfuscation process does not introduce significant noise in the system. As the size and complexity of the original circuit increase, the random circuit may require a larger number of gates for complete functional corruption, which can potentially decrease the fidelity of the circuit after deobfuscation.

\section{Deobfuscation Process and Overhead Analysis}
\label{sec:deobfus}
\subsection{Deobfuscation Process} 
\begin{figure}[t]
    \includegraphics[width=0.45\textwidth]{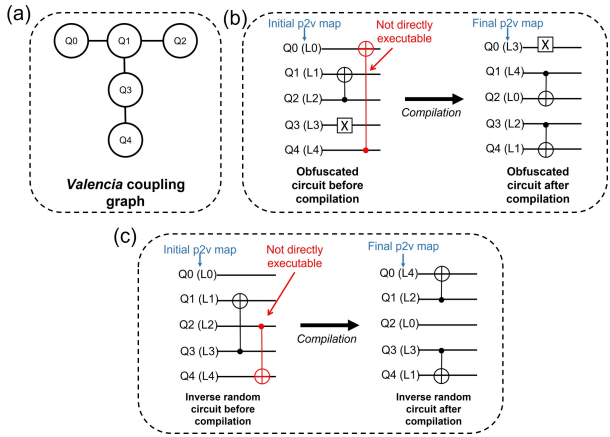}
    \centering
    \caption{(a) Coupling graph of $ibmq\_valencia$. (b) Example of physical-to-virtual (p2V) mapping of the obfuscated circuit before and after compilation. (c)  Example of p2V mapping of the inverse random circuit before and after compilation.}
    \label{fig:p2v}
    \vspace{-.31cm}
\end{figure}

After receiving the compiled circuits from the untrusted compiler, the user can apply a deobfuscation technique to restore the original functionality of the circuit. Our proposed deobfuscation technique involves several steps. Firstly, the user generates the inverse of the random circuit which was inserted into the original circuit. Next, he/she compiles the inverse circuit using either a trusted or untrusted compiler. Finally, the user inserts the compiled inverse random circuit into the compiled obfuscated circuit at the location where the random circuit was inserted before compilation. This ensures that the inverse circuit undoes the operations performed by the random circuit, effectively restoring the original functionality of the circuit. For the integrity of computation, it is crucial to maintain the continuity of physical-to-virtual (p2v) qubit mapping when concatenating the compiled inverse circuit with the compiled obfuscated circuit. Fig.\ref{fig:p2v} illustrates this with an example where a 5-qubit obfuscated circuit (Fig. \ref{fig:p2v}(b)) is to be run on the Valencia architecture (Fig. \ref{fig:p2v}(a)). We assume an initial p2v mapping of ${Q0 \to L0, Q1 \to L1, Q2 \to L2, Q3 \to L3, Q4 \to L4}$.  However, the gate \textit{CX L4, L0} cannot be directly executed with this mapping as there is no edge between \textit{Q4 (L4)} and \textit{Q0 (L0)} in Valencia’s coupling graph. This constraint is resolved by a SWAP operation during compilation, and the final p2v mapping is adjusted accordingly. Likewise, as shown in Fig. \ref{fig:p2v}(c), the initial p2v mapping of the inverse random circuit is also modified during compilation to comply with its coupling constraints. However, the final p2v mappings of these two circuits are not identical which poses a challenge during concatenation. The user can choose between two approaches to address this challenge: (a) feed the final mapping of the compiled obfuscated circuit as the initial mapping of the inverse circuit, or (b) add a SWAP layer between the compiled obfuscated circuit and compiled inverse circuit. Further details on these approaches can be found in \cite{saki2021split}.

One challenge to the proposed deobfuscation approach is the identification of the insertion location of the random circuit after compilation, particularly if it is inserted in the middle of the original circuit. Gates of the random circuit might get removed or combined with adjacent gates of the original circuit during optimization, making it difficult to determine the exact insertion location after compilation. To address this, the user can add a barrier on either side of the random circuit. This will leave a trace of the insertion location, without revealing it explicitly. The barriers are preserved during compilation. Therefore, post-compilation, the user can easily identify the insertion location and add the compiled inverse circuit next to the barrier on the side where the random circuit was inserted. It will still be difficult for the adversary to detect the random circuit and remove it since he/she must know (a) if the circuit is obfuscated, (b) the number of gates in the random circuit, and (c) accurately guess the side of the barrier where the random circuit is inserted. To make it more challenging for adversaries, additional barriers can be added around other gates of the obfuscated circuit. However, this may degrade the quality of optimization, as gates over the barriers are not optimized by the compiler.

\subsection{Overhead Analysis} 

\begin{figure}[t]
    \includegraphics[width=0.45\textwidth]{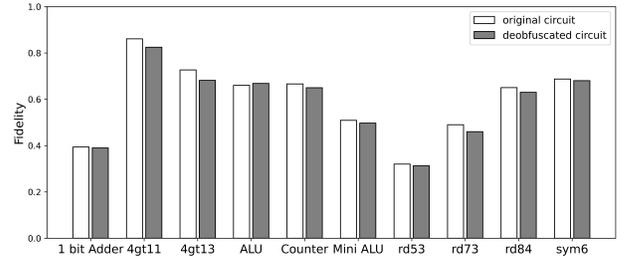}
    \centering
    \caption{Fidelity of original circuit and deobfuscated circuit for 10 benchmark circuits.}
    \label{fig:fidelity}
    \vspace{-.31cm}
\end{figure}

To quantify the overhead associated with the proposed obfuscation technique, we conduct a comparative analysis between the fidelity of the original quantum circuit and the deobfuscated circuit. Notably, the fidelity of a quantum circuit is defined as the ratio of the number of times the correct outcome is obtained out of the total number of shots. We use this metric to assess how closely the experimental outcomes of the deobfuscated circuit match the ideal expected outcomes. By comparing it with the fidelity of the original circuit, we can determine if the obfuscation and deobfuscation process introduces any significant differences in the circuit's performance. \\
We analyze the fidelity of 10 benchmark quantum circuits from the RevLib repository. 
We conduct experiments with 100 random and 100 refined random circuits, inserting them at three different locations of the benchmark circuits namely, front, middle, and back. 
Fig. \ref{fig:fidelity} shows the results for one such case in which we insert the refined random circuits at the back of the original circuit and employ the deobfuscation technique after compilation. The white bars represent the fidelity values of the original circuit, while the gray bars represent the average fidelity values of the 100 deobfuscated circuits. The results demonstrate that the proposed technique has a negligible (on average, less than 1\% for small-scale circuits and 3\% for larger and more complex circuits) impact on the fidelity of the circuit. Therefore, the proposed obfuscation and deobfuscation method restores the functionality of the original circuit with minimal degradation in fidelity but at higher security guarantees.

\section{Reverse Engineering Process and RE Effort Analysis}
\label{sec:reverse}

\begin{figure}[b]
    \includegraphics[width=0.45\textwidth]{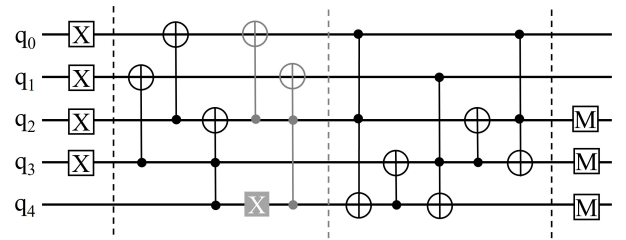}
    \centering
    \caption{Refined random circuit inserted in the middle of the counter circuit for obfuscation. Black gates and barriers belong to the original circuit, and the gray barrier and gates are part of the random circuit.}
    \label{fig:counter}
    \vspace{-.31cm}
\end{figure}

In this section, we analyze the effort required for an adversary to Reverse Engineer (RE) the original circuit from the obfuscated circuit. 

\subsection{Adversarial Assumption}
For this analysis, we consider a scenario where a user has sent an obfuscated circuit to an untrusted compiler after corrupting its functionality using our proposed technique. Similar to the adversary capability outlined in Section \ref{sec:threat_model}B, we assume that the rouge adversary is aware that the circuit he/she has obtained from the user has been obfuscated using the random circuit-based obfuscation technique. The objective of the adversary is to remove the random circuit and extract the original circuit with the correct functionality. However, the adversary cannot validate his guess in the absence of an oracle model. We also assume that the adversary can guess the location of obfuscation by observing the barrier. Furthermore, we assume that adversary is knowledgeable about quantum security fundamentals and knows that user may maximize metrics such as, TVD between original and obfuscated circuit for functional corruption. However, he/she may be unaware of the number of gates in the random circuit.

\subsection{RE Approach}

The adversary can adopt a pruning approach, where he/she systematically removes gates from each side of the barrier and compares the output of the pruned circuit with the output of the obfuscated circuit using the TVD metric. A high TVD value indicates greater dissimilarity between the outputs of the obfuscated and pruned circuits. The adversary anticipates that the original circuit will have a high TVD with the obfuscated circuit since the user has employed obfuscation techniques to corrupt the functionality. As a result, when the adversary obtains a low TVD from a pruned circuit, it indicates a closer similarity to the obfuscated circuit, and that choice can be discarded. Thus, the pruning approach can help the adversary in narrowing down the options by eliminating the incorrect choices.

To begin, we conduct experiments on a benchmark circuit (counter) and illustrate our findings. Fig. \ref{fig:counter} demonstrates the obfuscated counter circuit in which we have inserted a barrier in the middle and added a random circuit with three gates to the left side of the barrier. The left side of the barrier contains 6 gates, allowing the adversary to remove 1, 2, 3, 4, or 5 gates from that side in an attempt to extract the original circuit. Similarly, there are 5 gates on the right side of the barrier, giving the adversary 4 choices to make. However, the adversary cannot verify his/her guess as there is no oracle model available to confirm the correctness of the extracted circuit. It should be noted that this brute-force approach of trying different gate removal combinations is quite time-consuming. To be specific, the number of attempts required to extract the original circuit is \textit{n-2}, where \textit{n} represents the total number of gates in the obfuscated circuit.

\begin{table}[t]
\caption{TVD obtained after removing gates from the obfuscated counter circuit}
\begin{center}
\begin{tabular}{|c|cc|cc|}
\hline
\multirow{2}{*}{\textbf{\begin{tabular}[c]{@{}c@{}}No. of gates\\  removed\end{tabular}}} & \multicolumn{2}{c|}{\textbf{Removed from left}}                             & \multicolumn{2}{c|}{\textbf{Removed from right}}                            \\ \cline{2-5} 
                                                                                         & \multicolumn{1}{c|}{\textit{\textbf{TVD}}} & \textit{\textbf{Correct/Incorrect}} & \multicolumn{1}{c|}{\textit{\textbf{TVD}}} & \textit{\textbf{Correct/Incorrect}} \\ \hline
1                                                                                        & \multicolumn{1}{c|}{0.95}                  & Incorrect                            & \multicolumn{1}{c|}{0.97}                  & Incorrect                            \\ \hline
2                                                                                        & \multicolumn{1}{c|}{1.00}                  & Incorrect                            & \multicolumn{1}{c|}{1.10}                  & Correct                              \\ \hline
3                                                                                        & \multicolumn{1}{c|}{1.25}                  & Correct                              & \multicolumn{1}{c|}{0.31}                  & Incorrect                            \\ \hline
4                                                                                        & \multicolumn{1}{c|}{1.24}                  & Correct                              & \multicolumn{1}{c|}{1.03}                  & Incorrect                            \\ \hline
5                                                                                        & \multicolumn{1}{c|}{1.09}                  & Incorrect                            & \multicolumn{1}{c|}{}                      &                                      \\ \hline
\end{tabular}
\label{tab1}
\vspace{-.31cm}
\end{center}
\end{table}

Next, the adversary 
starts to remove one gate from each side, and then gradually increases the number of gates removed up to five on each side. The TVD is then calculated for each possible configuration of the pruned circuit. Table \ref{tab1} displays the results for all possible choices of gates removed from each side of the barrier. 
The\textit{ TVD} subcolumn shows the calculated TVD values for each choice, which quantifies the difference between the output of the pruned circuit and the obfuscated circuit. The \textit{Correct/Incorrect} subcolumn indicates whether the output of the pruned circuit is correct or incorrect compared to the expected output. Interestingly, our experiments reveal that the high TVD values are not always indicative of the correct output. Some configurations with high TVD values have resulted in incorrect outputs, which would further confuse the adversary. This demonstrates that the addition of the random circuit adds complexity to the RE process, making it challenging for the adversary to accurately identify the original circuit. However, upon closer examination of Table \ref{tab1}, we observe that for one of the pruned circuits, the adversary obtained a relatively small TVD value of 0.31. This indicates that the output of this pruned circuit is similar to the output of the obfuscated circuit, and the adversary may choose to discard this choice, thereby narrowing down their options to some extent using the pruning approach.

\subsection{Results}

\begin{figure}[t]
    \includegraphics[width=0.45\textwidth]{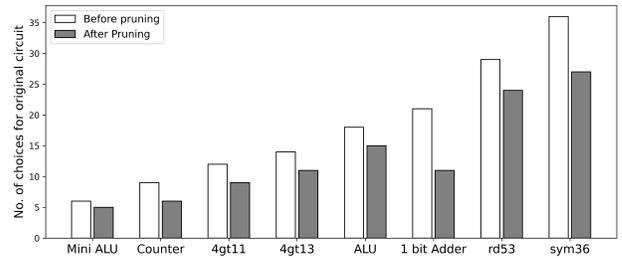}
    \centering
    \caption{Comparison between the number of potential choices for the original circuit before and after the adversary adopts the pruning approach to RE eight benchmark circuits.}
    \label{fig:pruning}
    \vspace{-.31cm}
\end{figure}

We extended the experiments to seven other benchmark circuits and present the results in Fig. \ref{fig:pruning}. The bar chart shows the number of possible choices before and after pruning, where each choice corresponds to a circuit configuration with a specific number of gates removed from each side of the barrier. Our findings show that the number of choices does not significantly reduce after pruning  indicating that the pruning approach is not much effective in aiding the adversary in identifying the original circuit from the obfuscated circuit. 

Further, we conducted the RE analysis for two other possible cases where the user inserts a random circuit at the front and back of the original circuit and obtained similar results. In these cases, the adversary can not guess the location of random circuit due to the absence of any additional barrier. So, he/she would have to try removing gates from both the front and the back until he/she was left with just one gate. This would further increase the number of potential circuits the adversary needed to check i.e., \textit{2*(n-1)} possible configurations.

Finally, it is important to note that even if the adversary manages to extract the original circuit through such brute force efforts, he/she will never be able to confirm his/her guess since there is no golden model available. Moreover, for this analysis, we took a pessimistic stand (from the user's perspective) and assumed the adversary is aware of the obfuscation and the obfuscation technique. Even for such a pessimistic case, the proposed obfuscation introduces a prohibitively large effort on the adversary's end. If we consider an optimistic case where the adversary does not know the type of obfuscation and the number of fake gates, the number of possible choices will significantly increase to \( \sum\limits_{k=1}^{n-1}{}_nC_k \) (\textit{n} = total no. of gates in the obfuscated circuit, \textit{k}= no. of gates removed), making the adversary's task much more challenging.



\section{Conclusion}
\label{sec:conclusion}

Quantum circuits can contain sensitive information, such as proprietary algorithms or financial data, which can be exposed to threats during compilation by an untrusted third party. This paper aims to secure the IP by inserting random circuits into the original quantum circuit before sending it to the untrusted compilers. We analyzed 18 benchmark circuits on IBM hardware architectures to evaluate the proposed idea and
achieved a high level of obfuscation, with a TVD of up to 1.92 while introducing negligible overhead and minimal degradation in fidelity (less than 1\% for small-scale circuits and below 3\% for large complex circuits). Additionally, our obfuscation method effectively defended against reverse engineering efforts which makes it a promising approach for securing sensitive IP embedded within quantum circuits.

\section*{Acknowledgements}
This work is supported in parts by NSF (CNS-1722557, CNS-2129675, CCF-2210963, CCF-1718474, OIA-2040667, DGE-1723687, DGE-1821766, and DGE-2113839), Intel’s gift and seed grants from Penn State ICDS and Huck Institute of the Life Sciences.

\bibliography{Ref}

\end{document}